\documentstyle[sprocl,psfig,epsfig,graphicx]{article}
\def\la{\langle}
\def\ra{\rangle}

\newcommand{\beas}{\begin{eqnarray*}}
\newcommand{\eeas}{\end{eqnarray*}}

\def\textmini#1{\textrm{\tiny #1}}

\title{ CLASSICAL ORDER PARAMETER DYNAMICS AND THE DECAY OF A METASTABLE
VACUUM STATE\footnote{Presented at Strong and Electroweak Matter
(SEWM2000), Marseille, France, 14-17 June, 2000. Based on an article
appeared in Phys. Rev. {\bf D62} (2000) 085013 written in collaboration    
with Szabolcs Bors\'anyi, Andr\'as Patk{\'o}s, J\'anos Polonyi.}}

\author{Zsolt Sz{\'e}p}
\address{Department of Atomic Physics, E{\"o}tv{\"o}s University \\
Budapest, Hungary}

\begin{document}
\maketitle

\abstracts{
Transition of the ground state of a classical $\Phi^4$ theory in $2+1$ 
dimensions is studied from a metastable state into the stable equilibrium. 
The transition occurs in the broken $Z_2$ symmetry phase and is triggered by 
a vanishingly  small amplitude homogeneous external field $h$. 
A phenomenological theory  is proposed in form of an effective equation of the order parameter which
quantitatively accounts for the decay of the  false vacuum. The large 
amplitude transition of the order parameter between the two minima displays 
characteristics reflecting dynamical aspects of the Maxwell construction.
}

The range of interest of the irreversible decay of a metastable vacuum state of 
finite energy density covers effects from cosmological phase transitions to 
instabilities observed in the mixed phase of first order phase transitions of
condensed matter systems \cite{boyanovsky}.

Whether the relevant mechanism for a first order phase transition is the
formation of bubbles of the new phase, as described by thermal
nucleation theory, or the gradual change of a large region of the sample, 
due to small amplitude spinodal instabilities described by spinodal
decomposition is also an intriguing question in heavy ion physics where the
actual expansion rate of the plasma may favour one or the other scenario
\cite{fraga}.

The conventional treatment of the decay of metastable states is based on the
nucleation theory but concurrent small amplitude spinodal instabilities are 
also present in the system. They are responsible for the flattening of the 
static effective potential (Maxwell cut) \cite{polonyi}. The clarification of their role
in the metastable $\to$ stable transition is the main
theme of the present investigation. 

\section{The model and the time-history of the order parameter (OP)}

We study numerically the dynamics of a classical $\Phi^4$ theory in $2+1$ 
dimensions governed by the discretized field equation of motion
$
\Phi_{\hat n}(t+a_t)+\Phi_{\hat n}(t-a_t)
  -2\Phi_{\hat n}(t)
  +a_t^2(-\Phi_{\hat n}+\Phi_{\hat n}^3-h)
  -{a_t^2\over a^2}\sum_i (\Phi_{\hat n+\hat i}(t)+
  \Phi_{\hat n-\hat i}(t)-2\Phi_{\hat n}(t))=0,
$
with initial conditions :
$ \dot\Phi_{\bf \hat n}(t=0)=0 $, $\Phi_{\bf \hat n}(t=0)=
		\Phi_0+\xi_{\hat n}\Phi_1$
where $\xi_{\hat n}$ is an evenly distributed white noise in the range
$(-1/2,1/2)$ (all quantities are expressed in units of mass).
The corresponding initial kinetic power spectrum is 
$E_k({\bf k})\sim\omega^2({\bf k})=-1+4(\sin^2({k_xa}/2)+\sin^2({k_ya}/2))/a^2$. 
We have chosen $\Phi_0=0.815$ and $\Phi_1=4/\sqrt{6}$ which
corresponds to a temperature value $T_i=0.57$ assuring that the system
is in the broken symmetry phase.
Our goal is to describe the evolution of the system only in terms of an
effective equation for the OP $\Phi(t)=\frac{1}{V}\sum_{\bf n}\Phi_{\bf n}(t)$.
\begin{figure}
\centerline{
\includegraphics[width=5.5cm]{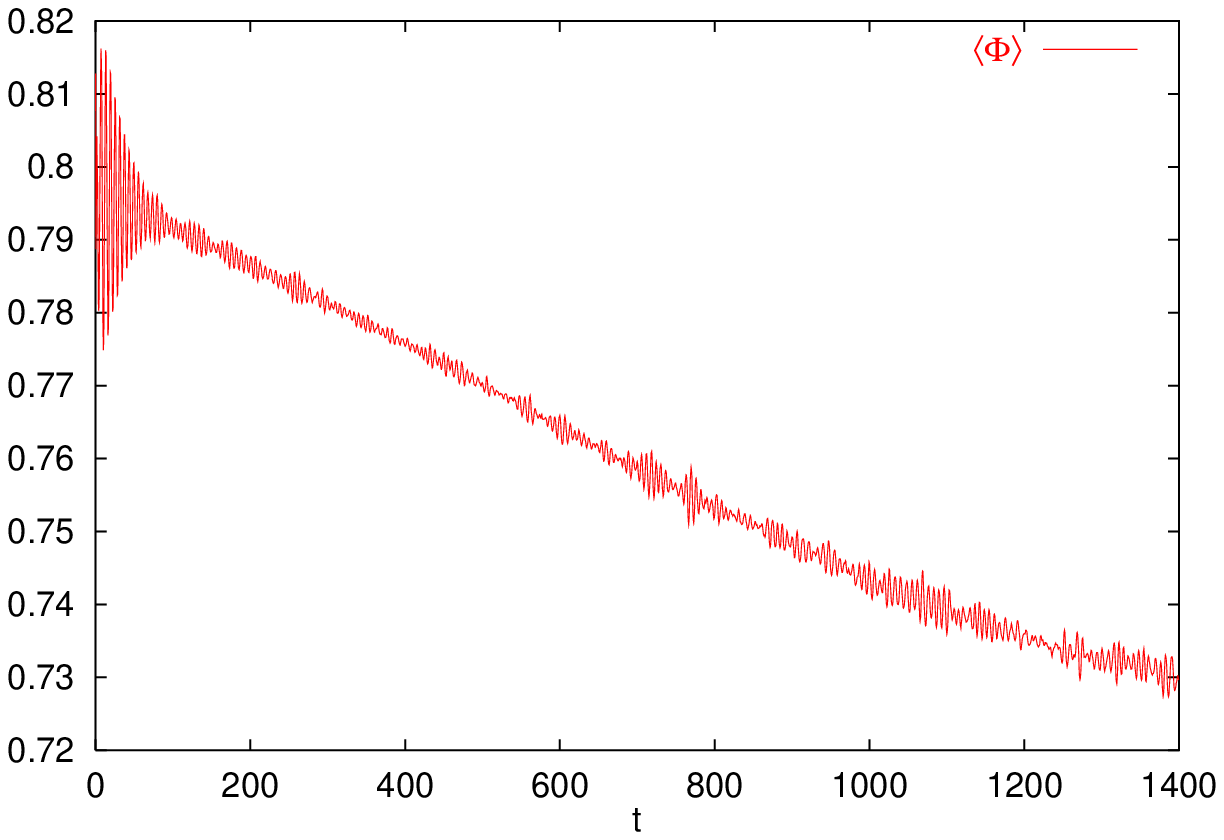}
\includegraphics[width=5.5cm]{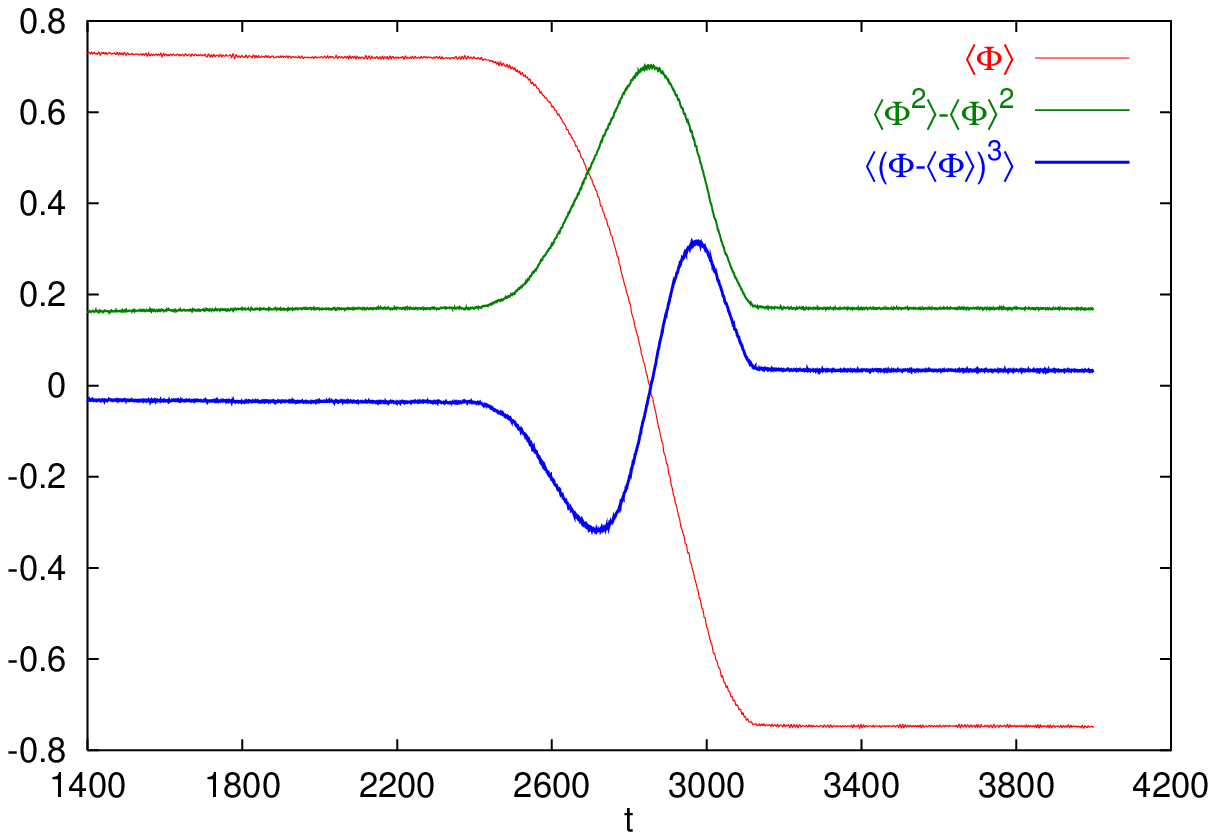}}
\caption{
Main qualitative features of the
$\Phi(t)$-trajectory for a $512 \times 512$ lattice and the external field 
$h=-0.04/\sqrt{6}$.}
\label{time_hist}
\end{figure}

Fig.~\ref{time_hist} shows a typical OP-history  together with the
later time history of the OP mean square (MS)-fluctuation 
($\langle\Phi^2\rangle -\langle\Phi\rangle^2$) and of its third moment 
($\langle (\Phi -\langle\Phi\rangle )^3\rangle $).
In general one can distinguish five qualitatively distinct parts of the 
OP-trajectory that starts with large amplitude damped oscillations
corresponding to the excitation of resonating modes, followed by a rather
slow relaxation to a metastable state characterised by 
$\langle\Phi\rangle\simeq0.72$. The (quasi)thermal motion in the
metastable state is followed by the 
metastable $\rightarrow$ stable
transition induced by the external field $h$, during which we can observe 
characteristic variations of the second and third moments. Quantitative 
interpretation of this variation will be presented in the following section. 
The last portion of the trajectory represents thermal motion in the true 
ground state.

\section{The effective OP-theory}
\begin{figure}
\centerline{  
\includegraphics[width=6cm]{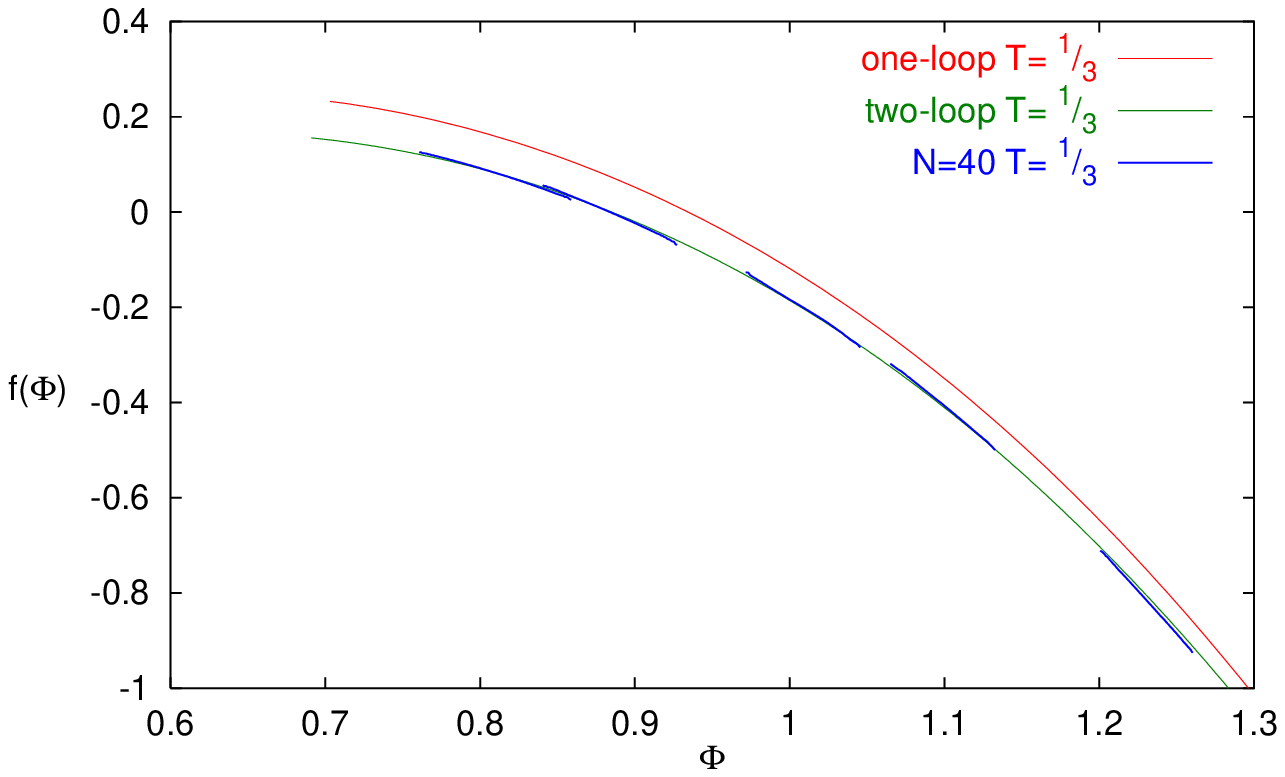}
\includegraphics[width=6cm]{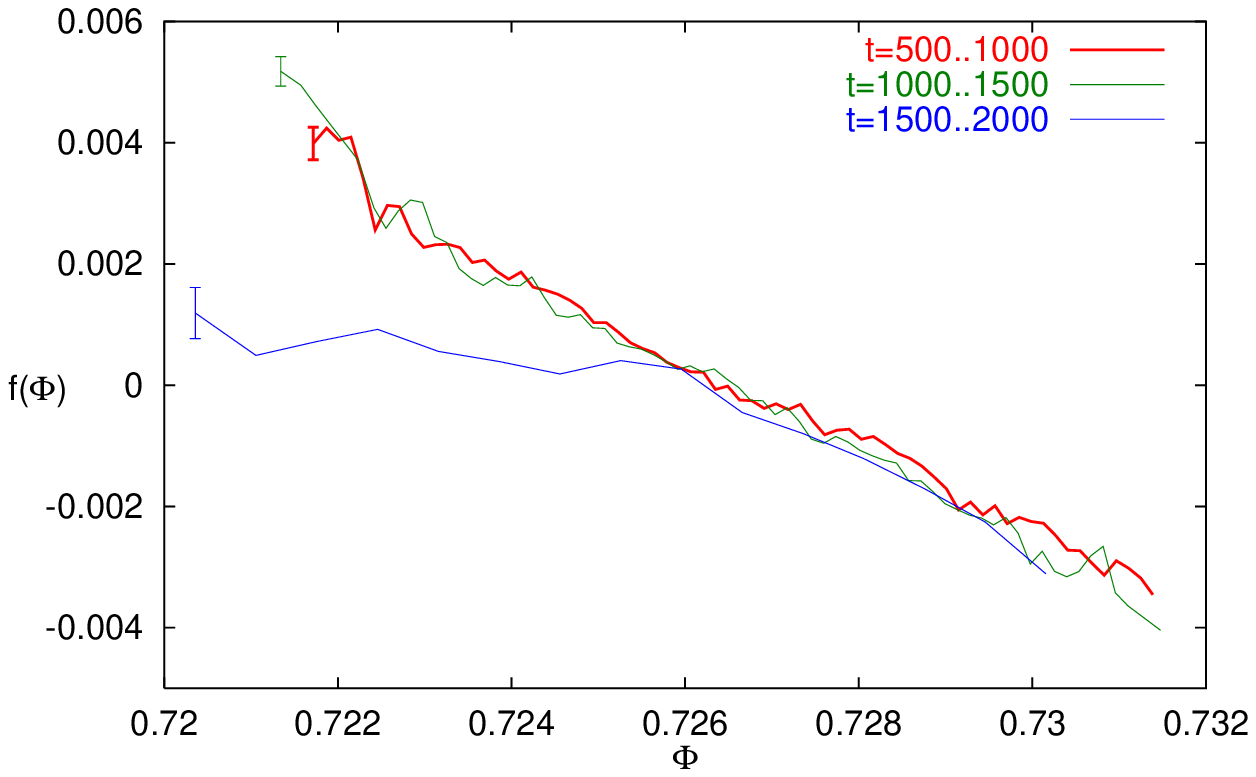}
}
\caption{Comparison of the force measured on a $40\times 40$ lattice with
te one calculated from a one-loop and a two-loop effective potential. On the
right hand one can see the bending down of the force prior the transition
time for a $100\times 100$ lattice and $h=-0.04/\sqrt{6}$.}
\label{force}
\end{figure}

For the description of (quasi)thermal motion near a (meta)stable point we
assume the validity of the ergodicity hypothesis for a system which
consist of a single degree of freedom, the OP of the system.
We describe its local time evolution by an effective Newton 
type equation:
$\ddot\Phi(t)+\eta(\Phi)\dot\Phi(t)+f(\Phi)=\zeta(t)$.
$\eta(\Phi)$,\,$f(\Phi)$,\,$\zeta(t)$ are obtained by a fitting procedure
attesting in this way also the presence of a term violating time-reversal
invariance $\eta(\Phi)$.  $\zeta(t)$ is the ``error'' of the best
global fit to the homogeneous equation at time $t$.
The force felt by the OP, $f(\Phi)$ agrees with the force derived from
the equilibrium two-loop 
effective potential as shown in Fig.~\ref{force}. To probe this agreement in
a relatively wide region one has to measure the force by shifting the center
of motion to different values of $\Phi$ by applying appropriate $h$ fields
to the system. The coefficient $\eta$ is well-defined and positive,
but its value depends quite substantially on the time resolution. The
analysis of the relationship of time averaging to the nonzero value of
$\eta$ is left for future investigations. 

\begin{figure}
\centerline{  
\includegraphics[width=5cm]{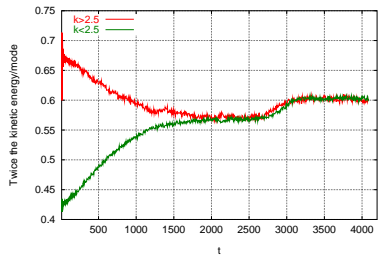}}
\caption{\label{k_hist}The time evolution of the kinetic energy content of  
the $|{\bf k}|>2.5$ and $|{\bf k}|<2.5$ regions averaged over the
corresponding $|\bf{k}|$-intervals.}
\label{thermal}
\end{figure}

The thermalization ``history'' of the system is shown in Fig.~\ref{thermal}.
The displayed temperature variations correspond to the kinetic energies
of the soft and hard modes.
During the time evolution they approach each other, manifesting relaxation
to a local thermal equilibrium. Because of our ``white noise'' 
initial condition the equilibration process is characterized by an energy
transfer towards the low $\bf\underline{k}$ modes.

The fluctuation moments depicted in Fig.~\ref{time_hist} along with the
measured force shown on the right of Fig.~\ref{force} tell about 
how the transition proceeds. The increased values of the moments indicate
the enhanced importance of the soft interactions. 
Preceding directly the
transition towards the direction of negative $\Phi$ values, the fitted force 
bends down and its average becomes a small positive constant. The
vanishing of the force implies the flatness of the effective potential along
the motion of the OP (the mode ${\bf k}=0$ in momentum space) indicating
the dynamical realization of the Maxwell-cut.

Taking into account the existence of a mixed phase during the transition 
period we can construct a model that reproduces exactly the shape of the 
moments and shows the vanishing of the force felt by OP.
Upon space averaging the microscopic field equation of motion 
of section 1 we find the following equation of motion for the OP:
$0=\ddot\Phi -\Phi+\Phi^3+3\overline{\varphi^2}^V \Phi
+\overline{\varphi^3}^V-h\equiv\ddot\Phi+dV_{inst}/d\Phi$
where the symbol $\overline{\varphi^n}^V$ means the space average of
$\varphi^n$, ($\overline{\varphi}^V=0, \Phi (t,{\bf x})=\Phi (t)+   
\varphi ({\bf x},t)$).
The instant potential $V_{inst}$, which is a fluctuating quantity,
contains as a {\it deterministic} piece the sum of the tree-level  
potential and of the slowly varying part of the second and third moments
$\overline{\varphi^n}^V(t)=(\overline{\varphi^n}^V)_{det}+\zeta_n(t)$,
\,$n=2,3$.
We assume that the space can be splitted into sharp domains 
(neglecting the thickness of the walls in between), where the field
is the sum of the constant background values 
$\Phi_{0\pm}$ and the fluctuations $\varphi_{\pm}$ around it,
$\Phi_\pm({\bf x},t)=\Phi_{0\pm}+\varphi_\pm({\bf x},t).$
Based on the smooth evolution of the temperature as displayed in 
Fig.~\ref{thermal} we assume also local equilibrium in both phases.

The actual value of the OP is determined by the surface
ratio $p(t)$ occupied by the stable phase:
${\Phi(t)}=p(t)\Phi_{0-}+(1-p(t))\Phi_{0+}.$
Simple calculation yields

$(\overline{\varphi^2}^V )_{det}(t)=
\frac{\Phi_{0+}-\Phi(t)}{\Phi_{0+}-\Phi_{0-}}
\left(\overline{\Phi_-^2({\bf x},t)}^V-\overline{\Phi_+^2({\bf
x},t)}^V\right)
+\Phi_{0+}^2
-\Phi^2(t)+\overline{{\varphi_{+}}^2}^V,$

$(\overline{\varphi^3}^V)_{det}(t)=
\frac{\Phi_{0+}-\Phi(t)}{\Phi_{0+}-\Phi_{0-}}
\left(\overline{\Phi_-^3({\bf x},t)}^V-\overline{\Phi_+^3({\bf
x},t)}^V\right)+
\overline{\Phi_+^3({\bf x},t)}^V\\
\hspace*{5cm}-3\Phi(t)(\overline{\varphi^2}^V)_{det}(t)-
\Phi^3(t)$.

\noindent
If one takes the values of $\Phi_{0\pm},\overline{ \varphi^n_{\pm}}^V$ 
from the respective equilibria    
a quite accurate description of the shape of the
two fluctuation moments arises in the whole transition region and its close
neighbourhood using the measured $\Phi(t)$ to parametrize their $t$-dependence. 
Substituting the expressions of the moments into the equation of the OP
one finds for the deterministic part of the force,
$f\left(\Phi\right)=
\frac{\overline{\Phi_+^3({\bf x},t)}^V-
\overline{\Phi_-^3({\bf x},t)}^V}{\Phi_{0+}-\Phi_{0-}}
\Phi-\Phi+\frac{\Phi_{0+}\overline{\Phi_-^3({\bf x},t)}^V-
\Phi_{0-}\overline{\Phi_+^3({\bf x},t)}^V}{\Phi_{0+}-\Phi_{0-}}-h.$
The average of the equations of motion in the respective equilibria,
$\la\Phi_\pm^3({\bf x},t)\ra-\Phi_{0\pm}-h=0$
implies the vanishing of the deterministic force, when exploiting the 
equality of averaging over the volume and the statistical ensemble.   
The equation of motion for OP 
goes into
$\ddot\Phi(t)+\zeta_3(t)+3\zeta_2(t)\Phi(t)=0$ which reflects 
the dynamical realization of the Maxwell construction when
assuming the validity of local equilibrium in both phases.

\section{The nucleation picture}
The statistics of the release time, the time necessary for the system to 
escape from the metastable region by nucleating a    
growing bubble of the stable state, is of the form $P(t)\sim
\textrm{exp}(-t\Gamma(h)L^2)$ where $\Gamma(h)=A\,\textrm{exp}(-S_2(h)/T)$ is   
the nucleation rate and $L$ is the lattice size \cite{gleiser}. 
We extract the exponent $S_2$ by fitting the measured rate to the expression 
of $\Gamma(h)$ assuming the $h$-independence of $A$. Comparing with the 
bounce action of the nucleation theory we obtain: 
$S_{\textmini{thin wall}}=\frac{4\pi}{9}h T\sim15S_{\textmini{measured}}$.
Using the $T\ne 0$ effective potential instead of the bare one one can
achieve a much better agreement, but still a factor of 2 
discrepancy remains with the mean field approach.

\section{Nucleation vs. spinodal instabilities}
Our results offer a ``dualistic'' resolution of the competition between the
nucleation and the spinodal phase separation mechanisms in establishing the
true equilibrium. We find that the statistical
features of the decay of the false vacuum agree with the results of
thermal nucleation. Alternatively, the effective OP-theory displays the 
presence of soft modes and produces dynamically a Maxwell cut when the time
dependence of the transition trajectory is described in the effective OP
theory. During the transition the OP travels trough a narrow but flat
valley around the ${\bf k}=0$ mode along which we expect the effective 
potential to be flat. This flatness is reflected also in the decrease 
of the kink-like action $S_{measured}$ relative to $S_{thin wall}$.
The larger is the system the smaller is the 
external field which is able to produce the instability. 

\end{document}